\documentclass[reprint,superscriptaddress,aps,prl]{revtex4-2}

\usepackage[usenames,dvipsnames]{xcolor}
\usepackage[utf8]{inputenc} 
\usepackage{bm}
\usepackage{graphicx,psfrag}
\usepackage{booktabs}
\usepackage{amsmath,amsthm,amssymb,amsfonts}
\usepackage{mathtools}
\usepackage{physics}
\usepackage{calc}
\usepackage{makecell}
\usepackage{enumitem}
\usepackage{float}
\usepackage{natbib}
\usepackage{tikz}
\usepackage{soul,xcolor}
\usepackage[colorlinks=true,linkcolor=Blue,citecolor=Blue,urlcolor=Blue]{hyperref}

\usepackage{xfp}

\usepackage[caption=false]{subfig}
\usepackage[capitalise]{cleveref}

\newcommand{\YZ}{\color{black}}
\renewcommand{\P}{\mathbb{P}}
\DeclareMathOperator{\err}{err}

\begin{document}

\title{The size of the sync basin resolved}

\author{Pablo Groisman}
\email{pgroisma@dm.uba.ar}
\affiliation{Departamento de Matem\'atica, Facultad de Ciencias Exactas y Naturales, Universidad de Buenos Aires}
\affiliation{IMAS-UBA-CONICET, Buenos Aires, Argentina}
\affiliation{NYU-ECNU Institute of Mathematical Sciences at NYU Shanghai}

\author{Cecilia De Vita}
\affiliation{Departamento de Matem\'atica, Facultad de Ciencias Exactas y Naturales, Universidad de Buenos Aires}
\affiliation{IMAS-UBA-CONICET, Buenos Aires, Argentina}

\author{Juli\'an Fern\'andez Bonder}
\affiliation{Departamento de Matem\'atica, Facultad de Ciencias Exactas y Naturales, Universidad de Buenos Aires}
\affiliation{Instituto de C\'alculo UBA-CONICET, Buenos Aires, Argentina}

\author{Yuanzhao Zhang}
\email{yzhang@santafe.edu}
\affiliation{Santa Fe Institute, Santa Fe, NM, USA}

\begin{abstract} 
Sparsely coupled Kuramoto oscillators offer a fertile playground for exploring high-dimensional basins of attraction due to their simple yet multistable dynamics. For $n$ identical Kuramoto oscillators on cycle graphs, it is well known that the only attractors are twisted states, whose phases wind around the circle with a constant gap between neighboring oscillators ($\theta_j = 2\pi q j/n$). It was conjectured in 2006 that basin sizes of the twisted states scale as $e^{-kq^2}$ with the winding number $q$. Here, we provide new numerical and analytical evidence supporting the conjecture and uncover the dynamical mechanism behind the Gaussian scaling. The key idea is that, when starting with a random initial condition, the winding number of the solution stabilizes rapidly at $t \propto \log n$, before long-range correlation can develop among oscillators. This timescale separation allows us to calculate the winding number as a sum of weakly-dependent random variables, leading to a Central Limit Theorem derivation of the basin scaling.
\end{abstract}

\maketitle

Basins of attraction map initial conditions to attractors and are fundamental to the analysis of multistable dynamical systems \cite{milnor1985concept,aguirre2009fractal,ott2002chaos}.
Even simple equations can generate complicated basins \cite{crutchfield1988subbasins,crutchfield1988attractors,wiesenfeld1989attractor,kaneko1990clustering,kaneko1997dominance,timme2002prevalence,daza2024multistability},
as exemplified by Wada basins \cite{nusse1996basins}, fractal basin boundaries \cite{mcdonald1985fractal,grebogi1987chaos,motter2013doubly,zhang2023catch}, and riddled basins \cite{alexander1992riddled,sommerer1993physical,ott1993scaling,heagy1994experimental,ashwin1996attractor,zhang2020critical}.
Given the intricate and often high-dimensional nature of basins, it is perhaps not surprising that even the most basic question---how big are the basins---still holds plenty of mystery \cite{wiley2006size,xu2011direct,monteforte2012dynamic,menck2013basin,menck2014dead,zou2014basin,martiniani2016turning,martens2016basins,leng2016basin,mitra2017multiple,delabays2017size,schultz2017potentials,rakshit2017basin,martiniani2017numerical,belykh2019synchronization,townsend2020dense,andrzejak2020two,andrzejak2021chimeras,zhang2024deeper}.

One of the canonical 
systems for studying basins is 
Kuramoto oscillators on cycle graphs \cite{wiley2006size,delabays2017size,zhang2021basins,diaz2024exploring}:
\begin{equation}
	\dot{\theta}_j = \sin(\theta_{j+1}-\theta_j) + \sin(\theta_{j-1}-\theta_j), \quad j = 1, \ldots, n,
	\label{eq:kuramoto_nn}
\end{equation}
where $\theta_j \in [0, 2\pi)$ is the phase of oscillator $j$. 
Note that we assume periodic boundary conditions to close the ring.
The sync state $\theta_1=\dots=\theta_n$ is always an attractor of the system.
For $n>4$, \cref{eq:kuramoto_nn} has additional attractors in the form of phase-locked configurations with the oscillator's phases making $q$ full twists around the ring: $\theta_j = 2 \pi j q/n + c$. 
Here, $q$ is the winding number and $c$ is a constant.  
Such twisted states are stable if and only if $|q|<n/4$ \cite{delabays2017size}. 
By varying the network size $n$, one can easily change the number of attractors in the system.

In 2006, based on numerical evidence and heuristic arguments, Wiley, Strogatz, and Girvan~\cite{wiley2006size} conjectured that the basin size of $q$-twisted states follows a simple scaling law of $e^{-kq^2}$, where $k$ is some constant. 
The conjecture was later challenged in the literature. For instance, based on semi-analytical calculations, Ref.~\cite{delabays2017size} suggested that the correct scaling should be $e^{-k|q|}$.
More recently, there was additional evidence supporting the original Gaussian scaling based on the geometries of the basins~\cite{zhang2021basins}.
Because the basin size decreases rapidly with $q$, basins with $q>\sqrt{n}$ can be exceedingly difficult to sample and direct numerical simulations cannot conclusively resolve the debate.
It is thus important to establish the scaling relation through analytical means. 

In this Letter, we show that basin sizes of twisted states in \cref{eq:kuramoto_nn} scale as $e^{-kq^2}$.
We break the argument into three steps:
\begin{enumerate}
    \item \label{one} Show the existence of a region $\mathcal{I}$ that is flow-invariant under the dynamics of \cref{eq:kuramoto_nn} and the winding number does not change once the system enters $\mathcal{I}$.
    \item \label{three} Show that up to $t \propto \log n$, there is no long-range dependence between the oscillators. Consequently, we can apply the Central Limit Theorem (CLT) to establish that the winding number (given by a sum of the phases) is Gaussian distributed at these times.
    \item \label{two} Show that when starting from a random initial condition, the system enters the region $\mathcal I$ quickly at $t \propto \log n$. This bounds the time window for which the winding number can change. Since the CLT holds for the winding number when entering $\mathcal I$ and it remains invariant after, the Gaussian scaling must hold for the final winding number at $t \to \infty$.
\end{enumerate}



Before giving details on these steps, we provide some rationale behind our strategy. It is more convenient to work with the phase differences between consecutive oscillators rather than directly with the phases $\theta_j$. We consider the new variables $\eta_j= \theta_{j+1} - \theta_{j} \in (-\pi,\pi]$. For $j=n$ we define $\eta_n= \theta_1 - \theta_n$. It is important to note that we force $\eta_j$ to be in the interval $(-\pi,\pi]$. In these new variables, \cref{eq:kuramoto_nn} is equivalent to
\begin{align}
\label{eq:eta}
\dot \eta_j(t) = \sin \left(\eta_{j+1}\right) -2\sin \left(\eta_{j}\right) + \sin \left(\eta_{j-1}\right),
\end{align}
with the caveat that the equation has to be interpreted mod $(-\pi,\pi]$. 
With this convention, if all the phase differences $\eta_j \neq \pi$,
we can compute the winding number as
\begin{equation}
    q(t) =  \frac 1{2\pi}\sum_{j=1}^n \eta_j(t) = \left[\frac 1{2\pi}\sum_{j=1}^{n-1} \eta_j(t) \right],
    \label{eq:winding_number}
\end{equation}
where $[x]$ denotes the closest integer to $x$.

\begin{figure}[t]
\centering
\includegraphics[width=.99\columnwidth]{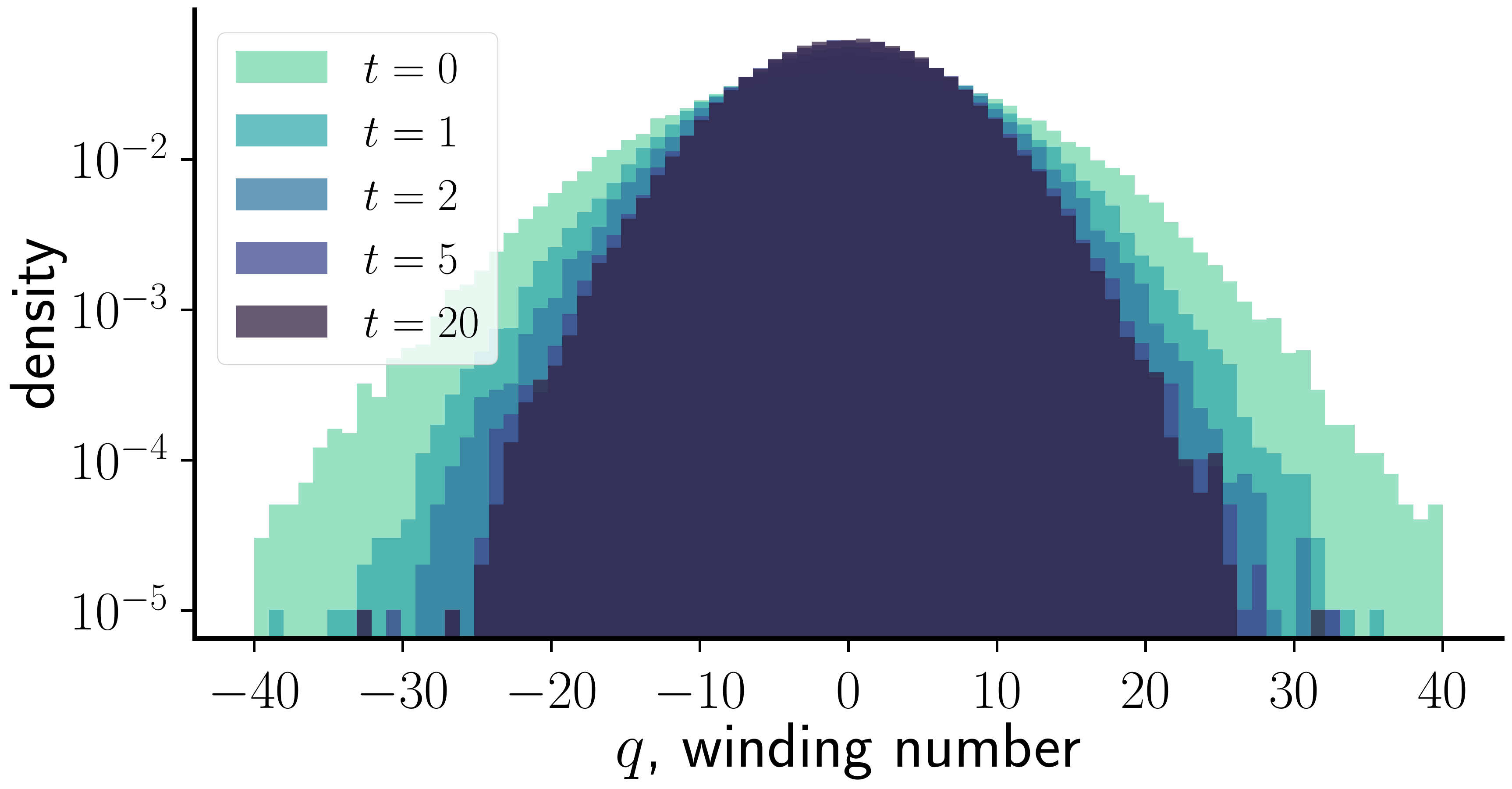}
\vspace{-4mm}
\caption{\YZ Distribution of the winding number $q$ at different times $t$ when starting from random initial conditions at $t=0$. The distribution remains Gaussian for all $t$, with the variance gradually decreasing during the early stage of the evolution but quickly converges to its final shape. Here, we set the number of oscillators $n=1280$ and each distribution is estimated from $10^5$ samples.}
\label{fig:q_distribution}
\end{figure}

Because the phase differences at $t=0$, $\eta_j(0)$, are independent random variables uniformly sampled from $(-\pi,\pi]$, by the CLT the winding number (their sum) follows a normal distribution when $n$ is large (\cref{fig:q_distribution}). 
The mean of the winding number is zero and its variance is $(n-1)/12$, since random variables uniformly distributed in $[-1/2, 1/2]$ have variance $1/12$. 
To obtain a well-defined distribution in the limit of $n\to\infty$, we simply need to normalize $q$ by $\sqrt n$.
Observe that as $t\to \infty$ we lose the independence (in fact for $t=\infty$ we have $\eta_i=\eta_j$ for every $i,j$) and hence at this time the Gaussian scaling can not be obtained as a consequence of the Central Limit Theorem.
Moreover, the winding number $q$ is not conserved by the dynamics.
So, how can we demonstrate that the distribution of $q$ would remain Gaussian as $t\to \infty$?

\begin{figure}[tb]
\centering
\includegraphics[width=.99\columnwidth]{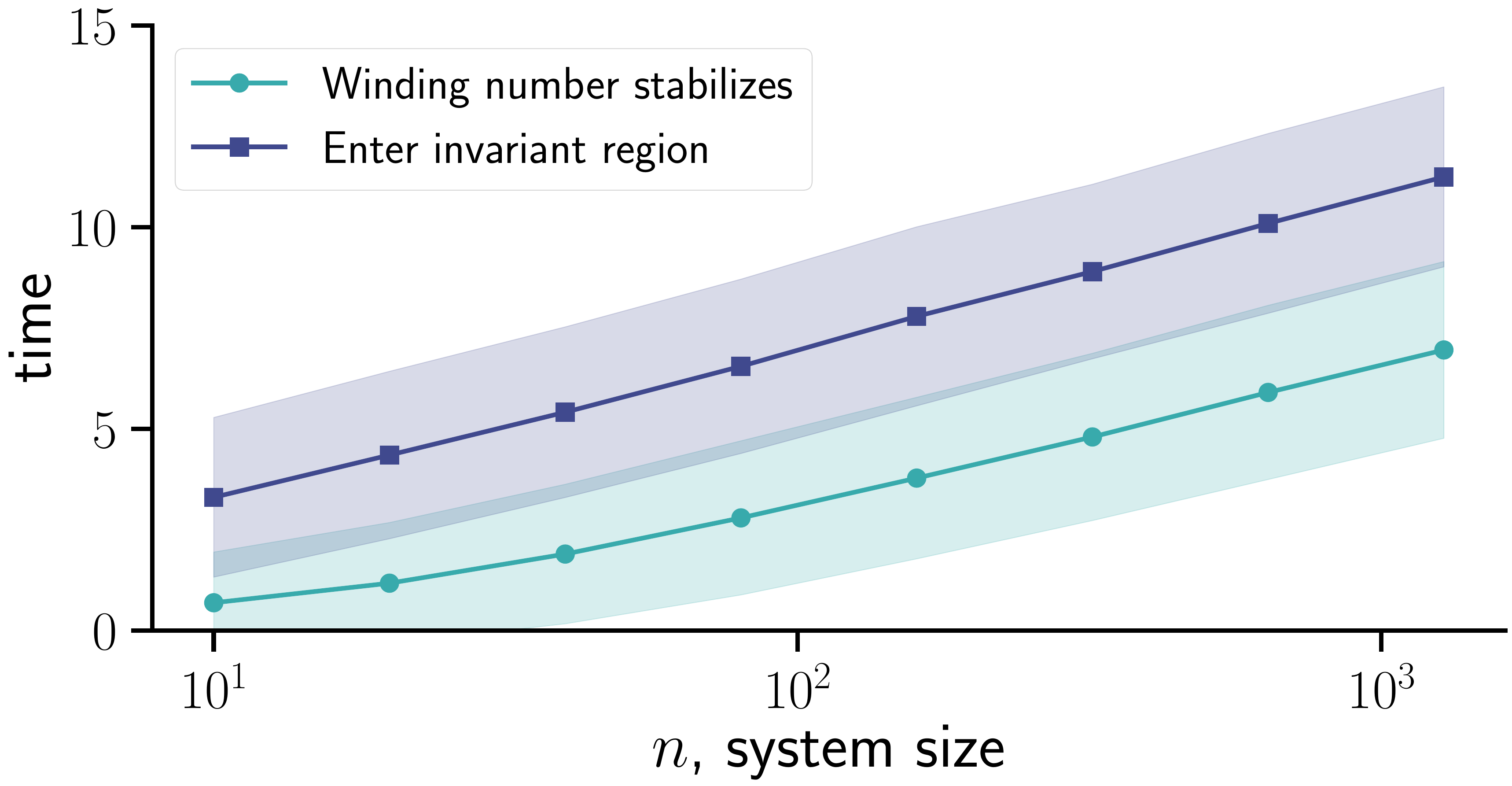}
\vspace{-4mm}
\caption{Time till the winding number stops changing ($t_s$) and time till the system enters the invariant region $\mathcal{I}$ ($t_e$) both scale as $\log n$.
This will be shown more rigorously in Step 3.
Each curve is averaged over $10^4$ trajectories starting from random initial conditions {\YZ and the shaded bands represent standard deviations}.
We always have $t_s \leq t_e$ for any individual trajectory, which is the point of Step 1.
}
\label{fig:t_vs_n}
\end{figure}

\begin{figure}[t]
\centering
\includegraphics[width=.99\columnwidth]{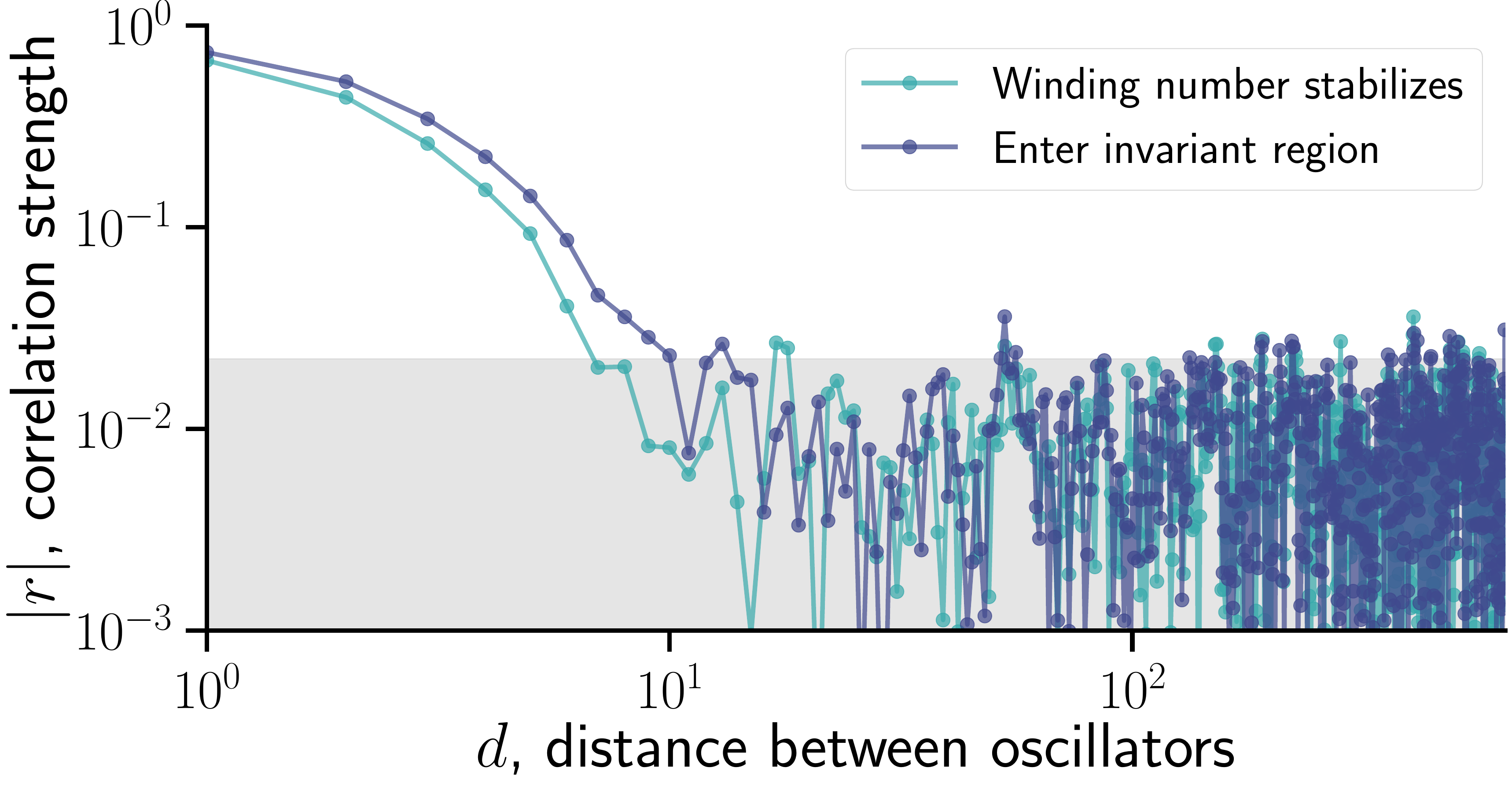}
\vspace{-4mm}
\caption{
No long-range correlation between oscillators can develop before the winding number stops changing.
Here, for $n=1280$ oscillators, we calculate the Pearson correlation $r$ between two oscillators that are distance $d$ apart at $t=t_s$ (winding number stabilized) and $t=t_e$ (entering the invariant region $\mathcal{I}$).
{\YZ The shaded area mark the expected $|r|$ for two random vectors whose entries are chosen uniformly and independently between $-\pi$ and $\pi$, $\mathbb E(|r|) = \frac{\sqrt{2}}{\sqrt{\pi}\sqrt{n-1}}\approx 0.0223$.}
The oscillators are essentially uncorrelated unless they are very close to each other ($d \leq 6$).
We show the lack of long-range correlation analytically in Step 2.
}\
\label{fig:iid}
\end{figure}

Numerically, we found that $q$ typically stabilizes very early on at a time $t_s$ and remains unchanged for $t > t_s$.
The green curve in \cref{fig:t_vs_n} shows that the average stabilization time $\langle t_s \rangle$ grows slowly with the system size as $\log n$.
The hope is that, at this early time, no long-range correlation has developed in the system and the CLT can still be applied to coarse-grained oscillator states. 
It is known that as long as the range of dependence is of order not larger than $n^{1/4}$, 
the CLT still holds \cite{francq2005central}.
Indeed, numerical evidence supports the no long-range dependence assumption (\cref{fig:iid}).
Later, in Step 2, we will explain this observation by utilizing the local coupling in \cref{eq:kuramoto_nn}.



We now proceed with the three steps of the argument.

{\bf \textit{Step 1.}}
Since it is not easy to estimate the stabilization time $t_s$ directly, as a first step, we would like to find a region in the phase space where $q$ would stay invariant. This would allow us to control the stabilization time $t_s$ by estimating the time $t_e$ it takes to enter the invariant region.
Since $t_s\leq t_e$, if we can show that for most initial conditions $t_e \propto \log n$, it would establish the desired bound $t_s \le \alpha\log n$, where $\alpha$ is a finite constant.
Indeed, \cref{fig:t_vs_n} provides numerical evidence that $\langle t_e \rangle \propto \log n$. We will also give analytical arguments for this in Step 3.

We denote $\bm{\eta}= (\eta_1, \dots, \eta_n)$ and consider the region $\mathcal I = \{\bm{\eta}\colon \eta_i \in (-\frac{\pi}{2},\frac{\pi}{2}) \text{ for all } i\}$.
We can establish its flow invariance through a maximum principle. 
Assume $\bm{\eta}(0) \in \mathcal I$ and let $t_0$ be the first time such that $\bm{\eta}(t_0) \in \partial \mathcal I$, the boundary of $\mathcal I$. 
Then, for some $i$ we have $\eta_i(t_0) \in \{-\pi/2, \pi/2\}$. 
Without loss of generality, we can assume $\eta_i(t_0) = \pi/2$. 
From \cref{eq:eta}, we have $\dot \eta_i(t_0) \le 0$ with strict inequality unless $\eta_{i-1}(t_0)=\eta_{i+1}(t_0)=\pi/2$. 
If we have strict inequality, we obtain a contradiction since $\eta_i$ needs to increase at $t_0$ to exit $\mathcal I$. 
If $\dot\eta_i(t_0)=0$, we have $\eta_{i-1}(t_0)=\eta_{i+1}(t_0) = \pi/2$, and proceeding inductively we obtain that either there is a node $j$ with $\eta_j(t_0) = \pi/2$ and $\dot \eta_j(t_0) < 0$, or $\eta_j(t_0) = \pi/2$ for every $j$. The state at which all the phase differences are $\pi/2$ is an unstable equilibrium and hence cannot be reached in finite time. We conclude that there is no such $t_0$ at which $\bm{\eta}$ can exit $\mathcal I$.

With a slightly more involved argument, we can show something stronger still. Let $\mathcal I_i = \{\bm{\eta}\colon \eta_i \in (-\frac{\pi}{2},\frac{\pi}{2})\}$, it is immediate that $\mathcal I = \cap_i \mathcal I_i$. We will show that in fact each $\mathcal I_i$ is invariant. 
In other words, once a phase difference enters $(-\frac{\pi}{2},\frac{\pi}{2})$, it will never leave.

We proceed with a perturbation argument. Instead of \cref{eq:eta}, consider the equations
\begin{equation*}
\label{eq:eta.epsilon}
\dot \eta^\varepsilon_i(t) = \sin (\eta^\varepsilon_{i+1}) -(2+\varepsilon)\sin (\eta^\varepsilon_{i}) + \sin (\eta^\varepsilon_{i-1}).
\end{equation*}
Assume $\eta_i^\varepsilon(0) \in \mathcal I_i$ and that at some finite time $t_0$ we have $\eta_i^\varepsilon(t_0)=\pi/2$ for the first time. 
Since
\[
\dot \eta^\varepsilon_i(t_0) \le  2 -(2+\varepsilon) = -\varepsilon,
\]
we have a contradiction. Thus, for every $\varepsilon >0$, $\eta_i^\varepsilon$ cannot leave $\mathcal I_i$.
Now, assume that there is a time $t_0$ such that $|\eta_i(t_0)| > \pi/2$.
By continuity of the solution at finite time $t_0$ with respect to the ODE parameters (see \cite[Theorem 2 on page 84]{Perko}), we have $\eta_i^\varepsilon(t_0) \to \eta_i(t_0)$ as $\varepsilon \to 0$. 
But $\eta_i^\varepsilon(t_0) \in  (-\pi/2,\pi/2)$ for every $\varepsilon$. 
This is a contradiction.
Hence, we conclude that there is no such $t_0$ and that $\eta_i$ cannot leave $\mathcal I_i$ once inside.

Next, we establish the invariance of the winding number inside $\mathcal{I}$. Since for $\mathbf \eta(t)\in \mathcal{I}$ formula \eqref{eq:winding_number} holds, 
we have
\[
\frac{d}{dt} q(t) =  \frac{1}{2\pi} \sum_{i=1}^{n} \frac{d}{dt}\eta_i(t) = 0.
\]
To see this more intuitively, note that for the winding number (a discrete quantity) to change along a continuous flow, it can only happen when one of the phase differences $\eta_i$ crosses $\pi$ or $-\pi$, the boundary points on which $q$ becomes ill-defined.
Since $\mathcal{I}$ does not include any of the boundary points and is flow invariant, there can be no more change in $q$ along the flow once inside $\mathcal{I}$. 

{\bf \textit{Step 2.}} 
In this step, our goal is to establish that long-range correlations cannot develop in \cref{eq:kuramoto_nn} at $t \propto \log n$, paving the way for the use of CLT.
For continuous time, it is difficult to control the correlation between two oscillators that are far away from each other. \textcolor{black}{This is because, although the dependence is small for oscillators that are far apart, all of them influcence each other at any positive time $t>0$. If we discretize time, in each time step each oscillator is only influenced by the value of the two neighboring oscillators at the previous time.}
So, we consider an Euler discrete scheme $\bm \eta^h$ with time step $h$ that approximates \cref{eq:eta},
\[
\eta_i^h(t_{k+1}) = \eta_i^h(t_{k}) + hG(\bm \eta^h(t_k)),
\]
with $G({\bm \eta}^h(t_k))=\sin(\eta_{i+1}^h(t_{k})) - 2\sin(\eta_i^h(t_{k})) + \sin(\eta_{i-1}^h(t_{k})))$ and $t_k=kh$. 
For this discrete scheme, it is easy to see that the range of dependence of an oscillator increases by two (one on each side) at each time step. 
\textcolor{black}{After $k$ steps, an oscillator depends on the initial values of $2k+1$ oscillators (the $k$ oscillators to its left, itself, and the $k$ oscillators to its right). Hence, the range of dependence grows linearly with time.}
To reach time $t = C \log n$, we need $Ch^{-1}\log n$ time steps. 
At that moment, the range of dependence for each oscillator is at most $Ch^{-1}\log n$. 
We can apply CLT as long as the step size $h$ 
approaches $0$ not too fast as $n\to \infty$ ($Ch^{-1}\log n \le n^{\kappa}$ for $\kappa <1/4$ is sufficient) \cite{francq2005central}.

Let us call \textcolor{black}{$\err_k$} = $\max_{1\le i \le n}$ $|\eta_i^h(t_k) - \eta_i(t_k)|$. An error analysis similar to Ref.~\cite[Section 2.6]{MM} gives $\err_k \le kh^2$. 
For $k=Ch^{-1}\log n$, $\err_k \le Ch\log n$, so we can control the error to go to $0$ as long as $h$ decreases faster than $\log^{-1}n$.
\textcolor{black}{In other words, the range of dependence of the oscillators at $t \propto \log n$ grows no faster than $\log^{2}n$, which is within the bound of $n^{1/4}$ required by the CLT.}


To export the CLT to the continuous equation we need to control \textcolor{black}{$q^h - q$}: the difference between the winding number of the discrete approximation and the one of the continuous solution (divided by $\sqrt{n}$). To do that, observe that 
this difference arises from those $\eta_i$ that are close to $\pi$ in the approximation and close to $-\pi$ in the solution of the ODE or vice versa (i.e. they contribute $\pm 2\pi$ to the difference of the sum involved in the computation of the winding number). \textcolor{black}{Thus, we can write $q^h - q = \sum_{i=1}^n X_i$, where
\[
X_i = \begin{cases}
    1 & \text{if $\eta_i^h \approx -\pi$ and $\eta_i \approx \pi$} \\
    -1 & \text{if $\eta_i^h \approx \pi$ and $\eta_i \approx -\pi$} \\
    0 & \text{otherwise}
\end{cases}.
\]}
\textcolor{black}{The variable $X_i$ equals $1$ when $\eta_i$ lies in the interval $[\pi - \err_k, \pi)$. The probability that $X_i = 1$ can then be bounded by $C\err_k$, where $C$ is a constant that depends on the density of $\eta_i$. Then, by symmetry, these variables are centered and their variances are bounded by $C \err_k$.}

So, a CLT holds and we can approximate the total difference $q^h - q$ with a Gaussian variable with standard deviation $\sqrt n$ \textcolor{black}{$\sqrt{C \err_k}$}. When we divide by $\sqrt n$, we get an error of order \textcolor{black}{$\sqrt{C \err_k}$}. In other words, the difference between the winding number of the discrete approximation given by the Euler method and the one of the continuous solution, when divided by $\sqrt n$, is also of order at most \textcolor{black}{$\sqrt{C \err_k}$}. So, we get the same condition as before  ($h < \log^{-1}n$). In this regime, the limiting distribution of the winding number of the approximation and the one of \cref{eq:eta} (divided by $\sqrt n$), coincide.

\begin{figure}[tb]
\centering
\includegraphics[width=.99\columnwidth]{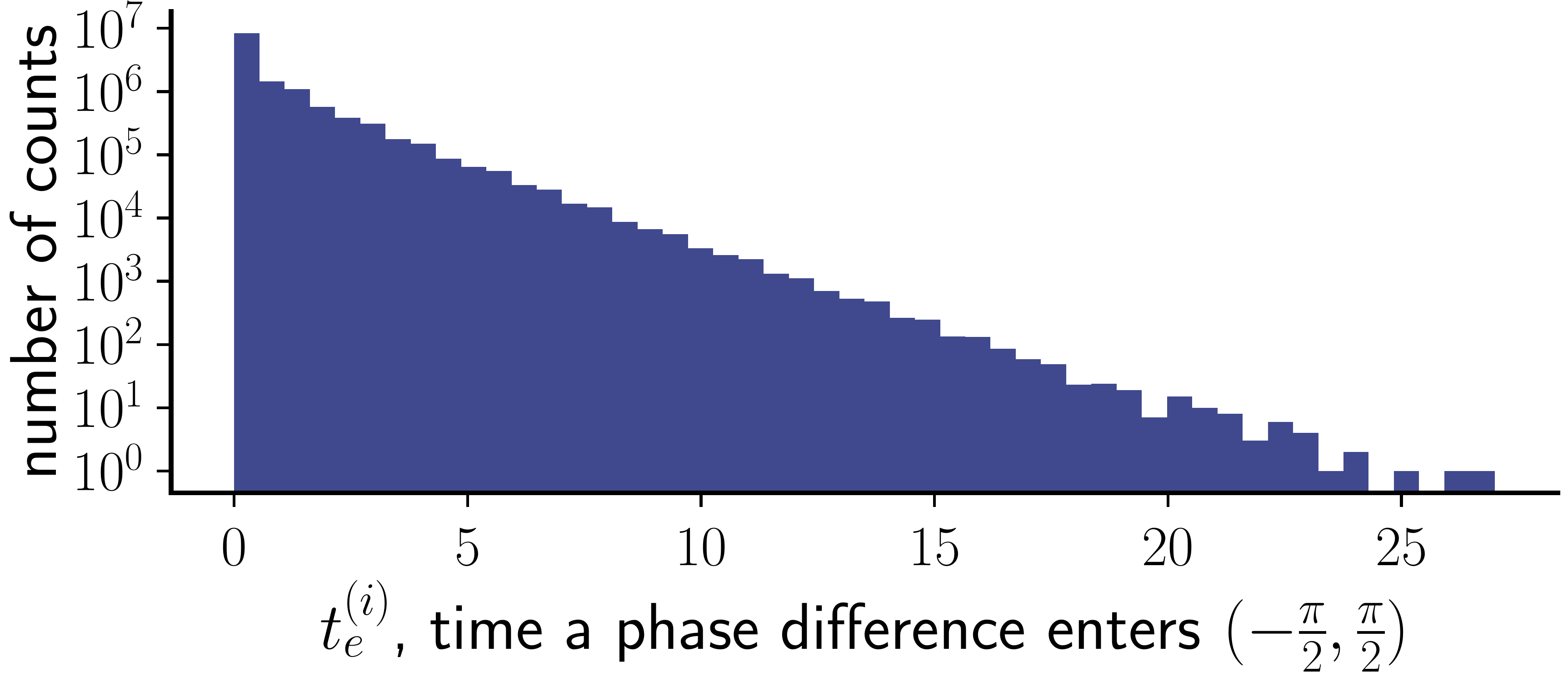}
\vspace{-4mm}
\caption{
Distribution of the times $t_e^{(i)}$ at which phase differences $\eta_i$ enter $(-\pi/2,\pi/2)$.
Half of the $\eta_i$ are already inside $(-\pi/2,\pi/2)$ for random initial conditions (thus the spike at $t=0$), while the nonzero entering times follow an exponential distribution. 
The data are collected from $10^4$ independent simulations of $n=1280$ Kuramoto oscillators from random initial conditions. The exponential distribution of $t_e^{(i)}$ is a key ingredient for Step 3.
}
\label{fig:Poisson}
\end{figure}

{\bf \textit{Step 3.}} 
The purpose of this step is to show that, starting from random initial conditions, the oscillators enter the invariant region $\mathcal I$ quickly at $t_e \propto \log n$, thus establishing the Gaussian scaling of winding numbers through the CLT.
If we look at each phase difference $\eta_i$ separately, we can define the time $t_e^{(i)}\geq 0$ at which it enters the interval $(-\pi/2,\pi/2)$ (from Step 1 we know that once entered, it will never leave). 
Due to symmetry, we know these times are identically distributed.
For the tail distribution function $F(t)=\P(t_e^{(i)} >t)$,
we have
\begin{multline*}
\P\left(\max_{1\le i \le n} t_e^{(i)} > a_n \right) = \P \left(\displaystyle \cup_i \{t_e^{(i)} >a_n\} \right) \\ \le n \P(t_e^{(i)} > a_n) = nF(a_n), 
\end{multline*}
\textcolor{black}{where we used the fact that the probability of the union of two events can be bounded from above by the sum of their probabilities.} To bound the entering time $t_e$ by $C\log n$, we want $nF(a_n) \to 0$ for $a_n= C\log n$.
This can be established if $F(t)$ has an exponential tail.
\textcolor{black}{In fact, it is enough to show that for any $C>0$ and some $\lambda >0$,  $F(t) \le e^{-\lambda t}$ for $t\le C \log n$, since in that case,  for $n$ large enough, the computation above reads,
\[
\P\left(\max_{1\le i \le n} t_e^{(i)} > a_n \right) \le n  e^{-\lambda C \log n} \to 0,
\]
as long as $C > 1/\lambda$.}
Combining with the fact that $\mathcal I_i$ is invariant, we have $t_e = \max_i t_e^{(i)} \leq C \log n$. 
Note that we do not need to assume independence among $t_e^{(i)}$ for this to hold.
In fact, strong correlation can quickly develop between $\eta_i$ and $\eta_j$ for neighboring $i$ and $j$, as can be seen in \cref{fig:iid}.
\Cref{fig:Poisson} provides numerical support by showing that $\P(t_e^{(i)})$ follows an exponential distribution, which implies that $F(t)=\P(t_e^{(i)} >t)$ also decays exponentially.

Below, we show why the distribution $F(t)$ has an exponential tail \textcolor{black}{by controlling how fast the energy decays in the system}.
\cref{eq:kuramoto_nn} is a gradient system, so its dynamics are fully determined by an energy function $E(\bm{\theta})$.
It is easy to see that
\begin{equation*}
    E_n(\bm{\theta}) = n - \frac{1}{2}\sum_{j=1}^n \left( \cos(\theta_{j+1}-\theta_j) + \cos(\theta_{j-1}-\theta_j) \right),
\end{equation*}
which can also be written in terms of 
$\bm{\eta}$ as 
\begin{equation*}
    E_n(\bm{\eta}) = n - \frac{1}{2}\sum_{j=1}^n \left( \cos(\eta_{j}) + \cos(\eta_{j-1}) \right).
\end{equation*}
\textcolor{black}{In the same way that we proved the CLT holds for $t\le C\log n $, we can show that in the same regime also the Law of Large Numbers hold for $g(\eta_i(t))$ for any continuous function $g$. In particular, we have that for $t\le C\log n$:}
\[
\frac 1 n E_n(\bm{\eta}) \to  1  - \mathbb E(\cos({\eta_1}(t))).
\]
Similarly, for the derivative of the energy we have \textcolor{black}{
\begin{align*}
\frac 1 n \dot E_n(\bm{\eta}) & = - \frac 1 n |\nabla E_n(\bm{\eta}(t))|^2\\
 & = - \frac 1 n \sum_{i=1}^n [\sin \left({\eta_i(t)}\right) - \sin \left(\eta_{i-1}(t)\right)]^2\\
  & = - \frac 1 n \sum_{i=1}^n [\sin^2 \left({\eta_i(t)}\right) + \sin^2 \left(\eta_{i-1}(t)\right)]\\
  & + \frac2n\sum_{i=1}^n\sin \left({\eta_i(t)}\right) \sin \left(\eta_{i-1}(t)\right)\\
 & \to -2\mathbb E(\sin^2(\eta_1(t))) + 2\mathbb E(\sin(\eta_1(t))\sin(\eta_2(t))).\\
\end{align*}
Next we notice that $\mathbb E(\sin(\eta_1(t)) \sin(\eta_2(t)))$ can be expressed as $\rm Corr(\sin(\eta_1(t)),\sin(\eta_2(t))) \mathbb E(\sin^2(\eta_1(t)))$. 
Up to time $C\log n$ the correlation between two consecutive oscillators has to be bounded away from one (otherwise it would contradict the fact that no long-range dependencies are developed shown in the previous step). So, for $t\le C\log n$,
\[
{\rm Corr}(\sin(\eta_1(t)),\sin(\eta_2(t))) \le c_1 < 1 \quad \text{and}
\]
\begin{align*}
\frac 1 n \dot E_n(\bm{\eta}) & \lesssim (-2 + 2c_1)\mathbb E(\sin^2(\eta_1(t))).\\
 & = -c_2\mathbb E(\sin^2(\eta_1(t))),
\end{align*}
with $c_2= 2(1-c_1)>0$}. Hence, if we show the existence of a positive constant $c$ such that $\mathbb E(\sin^2(\eta_1(t))) \ge c \mathbb E(1- \cos(\eta_1(t)))$ for times of order up to \textcolor{black}{$C\log n$}, we obtain that for such times and large $n$, with high probability \textcolor{black}{(i.e. probability converging to one as $n\to\infty$)},
\[
\dot E_n(t) \le -c E_n(t),
\]
which implies
\[
E_n(t) \le E_n(0)e^{-ct},
\]
and consequently $\mathbb E(1-\cos(\eta_i(t))) \le e^{-ct}$. Finally,
\begin{align*}
\P(t_e^{(i)}>t) & = \P(|\eta_i(t)|>\pi/2)\\ 
& = \P(1-\cos(\eta_i(t))>1)\\ 
& \le \mathbb E(1-\cos(\eta_i(t))) \\
&\le e^{-ct},
\end{align*}
where we used Markov inequality, \textcolor{black}{{which states that $\P(|X|>s)\le \mathbb E(|X|)/s$ for any random variable and $s>0$,}} to go from the second to the third line.

To show the existence of the positive constant $c$, observe that for any $\varepsilon>0$, we can choose $c>0$ such that $\sin^2(s) \ge c(1-\cos(s))$ for every $s\in(-\pi+\varepsilon, \pi-\varepsilon)$, with strict inequality except for $s=0$. Also observe that since $\mathcal I$ is invariant and contains all stable equilibria, we have $\P(|\eta_i(t)|<\pi/2) \to 1$ as $t\to\infty$ (and $\P(|\eta_i(t)|>\pi/2) \to 0$). So, the only thing that can prevent the existence of the constant $c$ is mass being lost at $|\eta(t)|=\pi$ at a slower rate than being gained at $\eta(t)=0$.
\textcolor{black}{Observe that $\tilde {\bm\eta}(t)= {\bm\eta}(-t) - \pi$ is also a solution of \eqref{eq:eta}. Hence, in the distribution of the random variable $\eta_i(t)$ (which does not depend on $i$)}, mass is grown at $0$ and lost at $\pi$ at the same rate. As a consequence,
\[
\inf_{t>0} \frac{\P(\varepsilon<|\eta_i(t)|<\pi-\varepsilon)}{\P(|\eta_i(t)|>\pi-\varepsilon)}>0.
\]
This is because if $\P(\varepsilon<|\eta_i(t)|<\pi-\varepsilon)$ converges to zero faster than $\P(|\eta_i(t)|> \pi -\varepsilon)$, that would mean that mass is growing at 0 faster than the rate at which is lost at $\pi$. Thus,
\[
\inf_{t>0} \frac{\mathbb E(\sin^2(\eta_i(t)))}{\mathbb E (1- \cos(\eta_i(t)))}=c>0,
\]
and we have, for \textcolor{black}{$t\le C \log n$}, $E_n(t) \le E_n(0)e^{-ct}$.

Now, combining all three steps, when the winding number is stabilized (at time $t \propto \log n$), we can establish the independence for phase differences $\eta_i$ and $\eta_j$ that are at distance $h^{-1}\log n = \log^{2} n$ from each other.
Because $\log^{2} n < n^{1/4}$, we can apply CLT to the phase differences
to obtain the Gaussian scaling.


In this Letter, we established
that the basin sizes in Kuramoto oscillators with nearest-neighbor coupling scales with winding number $q$ as $e^{-kq^2}$, contributing to a central debate on multistable dynamical systems spanning the past $20$ years.
Our results offer new insights into the dynamics of locally coupled Kuramoto oscillators (e.g., their winding number stabilizes early, before long-range correlations can develop), and the techniques developed here may also be applied to probe the basin sizes in other high-dimensional dynamical systems.
Future work has the opportunity to extend our results to more general network structures (e.g., ring networks with higher density \cite{wiley2006size}, signed networks \cite{diaz2024exploring}, non-regular networks \cite{kassabov2022global}, higher-order networks \cite{zhang2024deeper}, etc.) and dynamics beyond Kuramoto oscillators \cite{menck2014dead}.

\begin{acknowledgments}
We thank Steven Strogatz for the insightful discussions.
\end{acknowledgments}

\bibliography{bibli}

\end{document}